\documentclass[12pt]{article}
\usepackage[english]{babel}
\usepackage[cp1251]{inputenc}
\usepackage[a4paper, left=25mm, right=15mm, top=20mm, bottom=20mm]{geometry}
\usepackage{graphicx}
\usepackage{amssymb}
\usepackage{xcolor}
\usepackage{hyperref}
\hypersetup{pdfstartview=FitH, linkcolor=linkcolor,citecolor=citecolor,colorlinks=true} 
\usepackage{amsmath}
\usepackage{setspace}
\usepackage{bm}
\usepackage{cite}
\usepackage{url}
\usepackage[hang,flushmargin]{footmisc} 
\newcommand{\be}{\begin{equation}}
\newcommand{\ee}{\end{equation}}
\renewcommand{\baselinestretch}{1.25}
\date{}
\begin{document}
\definecolor{linkcolor}{HTML}{008000}
\definecolor{citecolor}{HTML}{4682B4}
\large
\title{\bf \LARGE Theoretical proof of the constancy of the speed of light in a vacuum}
\normalsize

\author{ D.N. Makarov$^{1,*}$ \\\\
$^{1}$Northern (Arctic) Federal University, Arkhangelsk, 163002, Russia\\
$^{*}$e-mail: makarovd0608@yandex.ru  }

\maketitle
\begin{abstract}
\begin{spacing}{1}
The constancy of the speed of light (the maximum velocity of interaction) is the second postulate of Albert Einstein's special theory of relativity. Currently, there is no correct theoretical proof of this constancy in all inertial frames of reference. This paper presents such a proof, demonstrating that quantum mechanics (quantum field theory) can only be formulated under the condition of the constancy of the speed of light in a vacuum. It has been established that this constancy is determined by the minimum energy of the particles. When this minimum is reached, two identical solutions emerge -- one with positive and one with negative energies. Thus, within the framework of classical physics, the existence of particles and antiparticles is demonstrated. It is shown that matter dominates over antimatter.
\end{spacing}
\end{abstract}

\section{Introduction}
The constancy of the speed of light is the second postulate of Albert Einstein's theory of relativity \cite{Einstein_1905}. The constancy of the speed of light means the constancy of the maximum speed of interaction, since the speed of light is equal to this speed.
The current state of physical science is based on this postulate, which has been experimentally confirmed to a very high degree of accuracy \cite{Guerrero_2025,Kostelecky_2011,Filippas_1964}. Of course, in experimental confirmation, the question always remains as to whether the speed of light will always remain constant, i.e., as the experimental accuracy increases. Of course, a definitive answer to this question is impossible, since any experiment has a limit to its accuracy. There is also the fundamental problem of substantiating the constancy of the speed of light in a vacuum: why is it constant and could this conceal any answers to other questions about our universe.

Currently, there is no correct theoretical proof of Albert Einstein's second postulate. A few statements about such proofs based on group theory can be found in the literature, where they are used to derive Lorentz transformations. Such proofs rely primarily on the works of Ignatovsky, Frank and Rothe \cite{Ignatowsky_1910,Ignatowsky_1911,Ignatowsky_1911_2,Ignatowsky_1911_3,Philipp_1911,Philipp_1912}, which has drawn justified criticism from the scientific community \cite{Pauli_1921,Miller_1981}. Without going into the details of these conclusions, here we will present a different approach based on invariant transformations and an analysis of the equations of motion, assuming that the speed of light is not a constant. Based on this, an expression will be derived linking the energy and momentum of a particle. By minimizing this energy, it will be possible to determine that the speed of light is a constant; in this case, the solutions will have two identical roots: one with positive and one with negative energies. We will also show that the constancy of the speed of light is a necessary condition for the existence of quantum mechanics (quantum field theory).

\section{Invariant transformations of space-time}
Let us show what invariant transformations should look like if the speed of light is not considered a constant value. For this purpose, we will consider space-time to be four-dimensional $\{x_{0},x_1,x_2,x_3\}$, where $x_{0}$ is the time component that is related to time $t$ through the proportionality coefficient $c$, i.e. $x_{0}=c t$, and $\{x_1,x_2,x_3\}$ are three spatial components. Next, to simplify the calculations, we will choose the particle's motion along one axis $x_1$. We will find the transformation law from one coordinate system $\{x^{'}_{0},x^{'}_1\}$ to another $\{x_{0},x_1\}$, i.e., the matrix $U$, where
\begin{eqnarray}
\begin{pmatrix}
  x^{'}_1\\\
  x^{'}_0
\end{pmatrix}=
U
\begin{pmatrix}
  x_1\\\
  x_0
\end{pmatrix},
~~~
U=\frac{1}{\sqrt{1-\delta^2}}\begin{pmatrix}
1& \delta\\\
  \delta& 1
\end{pmatrix} ,
\label{1}
\end{eqnarray}
where $\delta$ is some odd coefficient relative to the speed of movement $V$ of one reference system relative to another (the relative speed $V$), i.e. $\delta(V)=-\delta(-V)$. The matrix of Eq. (\ref{1}) is easily obtained by $U(V)=U^{-1}(-V)$ by replacing the relative velocity $V$ with $-V$ when passing from one reference frame to another. A detailed derivation of equation (\ref{1}) is given in many works and textbooks on special relativity. Since the function $\delta(V)$ is odd, it can be written as $\delta(V)=V A(|V|)$, where $A(|V|)$ is an even function with respect to $V$. In general, the matrix $U$ can be found by taking the differentials in Eq. (\ref{1}) of $x_0$ and $x_1$ and introducing the concept of velocity in the $K$ reference frame as $\frac{v}{c}=\frac{dx_1}{c dt}$ and in the $K^{'}$ reference frame as $\frac{v^{'}}{c^{'}}=\frac{dx^{'}_1}{c^{'} d t^{'}}$. In this case, the proportionality coefficients $c$ for $d x_0=cdt$ and $c^{'}$ for $d x^{'}_0=c^{'}dt^{'}$ are different. As a result, it is not difficult to obtain the relationship between the velocities of particles in different reference systems
\begin{eqnarray}
v^{'}=c^{'} \frac{v/c+V A}{1+ \frac{V A v}{c} }.
\label{2}
\end{eqnarray}
From Eq.(\ref{2}) it is clear that when $v=c$, we obtain $v^{'}=c^{'}$. We can also find particular values of the coefficient $A$. For example, when $v^{'}=0$, the relative velocity $V$ must be equal to the velocity $-v$, i.e. $v=-V$, then we obtain $A=1/c$. Similarly, when $v=0$, the relative velocity $V$ must be equal to the velocity $v$, i.e. $V=v$, then we obtain $A=1/c^{'}$. From this it is clear that $c$ and $c^{'}$ are the maximum possible velocities in the chosen reference frame, since from Eq.(\ref{1}) it is clear that $V A <1$. Thus, we identify the proportionality coefficient $c$ with the maximum possible speed, or, more simply, we will assume that this is the speed of light in the chosen reference frame. From Eq.(\ref{2}) one can find the coefficient $A$ in general form
\begin{eqnarray}
A =\frac{v^{'}/c^{'}-v/c}{V (1-\frac{v^{'}v}{c c^{'}} )}.
\label{3}
\end{eqnarray}
If we postulate that the speed of light is a constant, i.e. $c=c^{'}$, then from Eq.(\ref{3}) we can easily obtain that $A=1/c$ and we obtain from Eq.(\ref{1}) the well-known Lorentz transformation.

We add that transformation (\ref{1}) is more general than the Lorentz transformation. Transformation (\ref{1}) also corresponds to rotations in Minkowski space, i.e., rotations in the spacetime coordinate system, with the only difference being that the time axis $x_0=c t$ changes not only time but also the speed of light $c$ during rotations. Thus, we will henceforth assume that the speed of light $c$ depends on the speed $v$, i.e. $c=c(v)$. Also Eq.(\ref{1}) is invariant, i.e. $ds^2=(dx_0)^2-(dx_1)^2=(dx^{'}_0)^2-(dx^{'}_1)^2$. Further we will use the standard notation of four-vectors, i.e. $ds^2=dx^{i}dx_{i}$, where $x^{i}$ and $x_{i}$ are the contravariant and covariant components of the coordinates.

\section{Relationship between energy and momentum. Minimum energy}
Further calculations will be similar to those well known in the special theory of relativity, but taking into account that $c=c(v)$. To do this, we introduce an invariant action function $dS=\alpha ds= L(x,v,t)dt$, where $L=\alpha c(v)\sqrt{1-\left(\frac{v}{c(v)}\right)^2}$ is the Lagrangian of the system, and $\alpha$ is some coefficient that is sought in the same way as in the special theory of relativity. Since the space must be isotropic, the Lagrangian must depend only on the velocity module, which means $c=c(|v|)$, then $L=\alpha c(|v|)\sqrt{1-\left(\frac{v}{c(|v|)}\right)^2}$. The equation of motion with such a Lagrangian is similar to the previously known one, with the only difference being that the momentum will be in the form
\begin{eqnarray}
P =-\alpha \gamma \left(\frac{v}{c(|v|)}-\frac{d c(|v|)}{d v}\right),~ \gamma=\frac{1}{\sqrt{1-\left(\frac{v}{c(|v|)}\right)^2}}.
\label{4}
\end{eqnarray}
Similarly, we can do the same for the energy $E=P v-L$, we get
\begin{eqnarray}
E =-\alpha \gamma c(|v|) \left(1-\frac{v}{c(|v|)}\frac{d c(|v|)}{d v}\right).
\label{5}
\end{eqnarray}
It is well known that the quantity $(E/c)^2 - P^2$ is an invariant in special relativity. In our case, when $c=c(|v|)$, this is not at all obvious. To verify this, we need to introduce the energy-momentum four-vector $P^{i}=(P^0,P^1)$ and demonstrate its invariance, i.e. $P^{i}P_{i}=inv$. To do this, we introduce new space-time variables under the linear transformation $\varkappa^i=x^i-e^i_k x^k \beta \frac{v}{|v|}$, where $\beta$ is some constant, $e^i_k=1-\delta^i_k$, and $\delta^i_k$ is the Kronecker delta.
It is easy to see that the new variable $\varkappa^i$ is transformed when moving to another coordinate system in the same way as the variable $x^i$, i.e.
\begin{eqnarray}
\begin{pmatrix}
  \varkappa^{'}_1\\\
  \varkappa^{'}_0
\end{pmatrix}=
U
\begin{pmatrix}
  \varkappa_1\\\
  \varkappa_0
\end{pmatrix},
~~ \varkappa^i\varkappa_i={\varkappa^{'}}^i \varkappa^{'}_i=x^ix_i (1-\beta^2).
\label{6}
\end{eqnarray}
Since the parameters $\varkappa^i\varkappa_i$ and $x^ix_i$ must be positive values, and the coordinates are real numbers, then the parameter $|\beta| \leqslant 1$ and $\beta {\not =} \beta(v)$. Next, we introduce the four-velocity $u^i=\frac{d \varkappa^i}{d s}$. Similarly, we introduce the four-momentum $P^i=-\alpha u^i$. It is easy to see that we can introduce the invariant $P^iP_i=\alpha^2 \frac{d \varkappa^i}{d s} \frac{d \varkappa_i}{d s}=\alpha^2(1-\beta^2)$. Such a four-momentum $P^i$ will have the form $P^i$
\begin{eqnarray}
P^1 = -\alpha \gamma  \left(\frac{v}{c(|v|)}-\beta \frac{v}{|v|}\right),~~P^0 = -\alpha \gamma  \left(1-\frac{v}{c(|v|)}\beta \frac{v}{|v|} \right).
\label{7}
\end{eqnarray}
Comparing Eqs.(\ref{4}),(\ref{5}) with (\ref{7}) we come to the conclusion that the momenta and energies coincide at $\frac{d c(|v|)}{d v}=\beta \frac{v}{|v|}$. This means that from a four-dimensional momentum one can compose an invariant quantity $P^i P_i=inv$ only when the speed of light $c(|v|)=\beta |v|+c_0$, where $c_0$ is the speed of light at zero velocity. This result resembles (but does not fully correspond to) the well-known law of addition of velocities in classical mechanics; in any case, the speed of light is proportional to the speed $v$. At first glance, one could use another initial condition and find the parameter $\beta$, namely $P(v=0)=0$, while $\beta=0$, then $c=c_0$. Such an initial condition cannot be applied in the general case, since this case $P(v=0)=0$ initially assumes the constancy of the speed of light. It is worth noting that the momentum of the system may not be zero, even at $v=0$, the same is true for energy. The four-dimensional momentum $P^i$ has one interesting property, namely: as the velocity $v$ increases, $P^i$ also increases for any $|\beta| \leqslant 1$, i.e., the values of momentum and energy increase (assuming $\alpha<0$). The expressions for momentum and energy can be simplified, knowing the law of change in the speed of light $c(|v|)=\beta |v|+c_0$, and obtain
\begin{eqnarray}
P = -\alpha \gamma  \frac{v}{|v|} \left(\frac{|v|}{c(|v|)}-\beta \right),~~E = -\alpha \gamma  c_0.
\label{8}
\end{eqnarray}
Further, it is more convenient to use the impulse $p$, which is obtained when $P=p\frac{v}{|v|}$ ($p=p(|v|)$). Note that the impulse $p$ can also be two-valued, positive when $\frac{|v|}{c(|v|)}>\beta$ and negative when $\frac{|v|}{c(|v|)}<\beta$. 

The relationship between energy and momentum can also be obtained in the form
\begin{eqnarray}
\large
{\boxed{  E^2-(p c_0+\beta E)^2=\alpha^2c^2_0},~~  |\beta| \leqslant 1}.
\label{9}
\end{eqnarray}
We add that Eq.(\ref{9}) is asymmetric with respect to $E$ and $pc_0$. The nature of this asymmetry will be clarified below. Next, we express energy through momentum
\begin{eqnarray}
E=\frac{|\alpha| c_0}{\sqrt{1-\beta^2}}\left[\pm \sqrt{1+\left(\frac{\mathcal P}{\sqrt{1-\beta^2}}\right)^2}+\beta \frac{\mathcal P}{\sqrt{1-\beta^2}}\right], ~ \mathcal P = \frac{p}{|\alpha|}.
\label{10}
\end{eqnarray}
It can be seen that in Eq.(\ref{10}) two signs $\{\pm\}$ appear. These signs are not placed arbitrarily, i.e. either $\{+\}$ or $\{-\}$, but correspond to certain values of the parameter $\beta$ and the dimensionless momentum $\mathcal P$. This can be seen using the properties of energy and momentum described above, i.e., with increasing velocity, energy and momentum increase for all values of $|\beta| \leqslant 1$, which means that with increasing momentum, energy must increase. In mathematical notation, this means that $\frac{d E}{d p}=\frac{\partial E}{\partial v} \frac{\partial v}{\partial p}$, and since $\frac{\partial E}{\partial v} >0$ and $\frac{\partial p}{\partial v} >0$, then $\frac{d E}{d p}>0$. We add that in Eq.(\ref{8}) we put the sign $\{+\}$ before the energy, but this is a convention, since we can also put the sign $\{-\}$, the equations of motion will not change. If we put the sign $\{-\}$ before the energy in Eq.(\ref{8}), then with an increase in velocity the energy decreases, then in Eq.(\ref{10}) with an increase in momentum, the energy decreases. Taking this into account, the Eq.(\ref{10}) can be rewritten as
\begin{eqnarray}
E=\pm \frac{|\alpha| c_0}{\sqrt{1-\beta^2}}\left[ \sqrt{1+\left(\frac{\mathcal P}{\sqrt{1-\beta^2}}\right)^2}+|\beta| \frac{|\mathcal P|}{\sqrt{1-\beta^2}}\right].
\label{11}
\end{eqnarray}
In Eq.(\ref{11}) the signs of $\pm$ are not arbitrary, i.e. either $\{+\}$ or $\{-\}$, but correspond to specific values of the parameter $\beta$ and momentum $p$. For example, from the analysis of Eq.(\ref{10}) one can unambiguously state that the sign $\{-\}$ will appear for $\beta\in [-1,0)$ and $p>0$, but $\{-\}$ also appears for $p<0$, but for $\beta\in (0,1]$. From Eq. (\ref{11}), we can find the minimum value of the energy $E$ as a function of the parameter $\beta$. We add that $p$ is a dynamic variable and $\alpha$ is a constant related to the mass, so the only parameter to be minimized is $\beta$. As a result, the minimum energy is found when $\beta \to 0$. We clarify that the signs of $\pm$ in Eq. (\ref{11}) are arbitrary, i.e. $\{+\}$ can be replaced by $\{-\}$, but in this case, $\{-\}$ must be replaced by $\{+\}$, so when minimizing, we must seek the minimum for $|E|$.

An interesting fact is how the parameter $\beta$ tends to zero. If we tend to zero from the right $\beta \to 0+$, then we obtain an energy with the $\{+\}$ sign, but if we tend to zero from the left $\beta \to 0-$, we obtain an energy with the $\{-\}$ sign, but the absolute values of the energies will be the same. This means that there are two energy values at $\beta \to 0$, which corresponds to the well-known expression $E^2-p^2c^2_0=\alpha^2c^2_0$. We add that at $\beta \to 0$ the speed of light always becomes a constant, which corresponds to the proof of the constancy of the speed of light in any inertial reference frame. It can also be said that as $\beta\to 0$, a unique case arises where, for the same energy (in absolute value), there exist two separate solutions with the signs $\{+\}$ and $\{-\}$, which indicates the existence of two types of similar particles. Currently, we know that these are particles and antiparticles. We add that the well-known expression in special relativity $E=\pm \sqrt{\alpha^2c^2_0 + p^2c^2_0}$ does not imply the existence of two solutions at once, since the choice of signs is arbitrary and this can be interpreted as the choice of any sign to describe physics. In our solution, the energy is the same, but two signs determine the existence of two separate solutions.

An extremely interesting point is that the Eq.(\ref{9}) for non-zero values of $\beta$ is asymmetric. If we quantize this expression, i.e. $E\to {\hat H}=i\hbar \frac{\partial}{\partial t}$, and $p\to {\hat p}=-i\hbar \frac{\partial}{\partial x}$, we can see that a term appears in the differential equation $\frac{\partial}{\partial t} \frac{\partial}{\partial x}$ in the form of mixed derivatives. This means that it is impossible to separate the time and coordinate parts of the wave function separately, and hence there are no stationary solutions. This leads to the fact that for non-zero $\beta$ there is no such concept of particles as in quantum mechanics or quantum field theory. Strictly speaking, they are completely impossible to obtain. Ultimately, we can say that a minimum of energy, and therefore the constancy of the speed of light, ensures the existence of the quantum world as we know it.

\begin{figure}[h!]
\center{\includegraphics[width=0.95\linewidth]{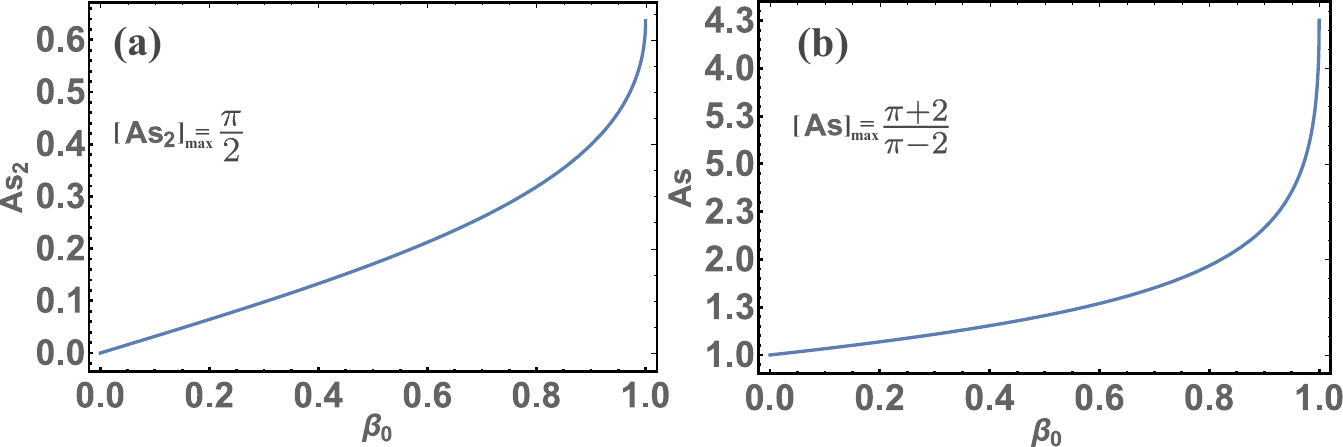}}
\caption[fig_1]{Figure (a) shows the symmetry parameter $As_2=\frac{N_{+}-N_{-}}{N_{+}+N_{-}}$ as a function of the parameter $\beta_0$. Figure (b) shows the same, but for the parameter $As=\frac{N_{+}}{N_{-}}$.}
\label{fig_1}
\end{figure}

If there is an asymmetry of solutions for nonzero $\beta$, then the number of particles and antiparticles, or more precisely, matter and antimatter, will be different. We can introduce an asymmetry parameter for such an estimate as $As=\frac{N_{+}}{N_{-}}$, where $N_{+}$ is the number of particles (or matter) with positive energy, and $N_{-}$ with negative energy. We assume that the number of particles is proportional to their energy $E$, then $N_{\pm}\sim \int_{\pm} E_{\pm}P(\beta)P(v) d \beta d v$, where $P(\beta)d \beta$ is the probability of detecting a particle in the element $d b$, similarly for $P(v)d v$ it is the probability of detecting a particle in the velocity element $d v$, integration is carried out in those intervals $\beta$ and $v$ where the energy corresponds to its sign. As a result, with an accuracy of up to an insignificant proportionality coefficient, one can imagine
\begin{eqnarray}
N_{+}={\int^1_0}P(\beta) \int^{\frac{1}{1-\beta}}_{\frac{\beta}{1-\beta^2}} \frac{ P(x=v/c_0)}{\sqrt{1-(\frac{x}{1+\beta x})^2}}  d x d \beta ,
\nonumber\\
N_{-}={\int^1_{0}}P(\beta)\int^{\frac{\beta}{1-\beta^2}}_{0} \frac{ P(x=v/c_0)}{\sqrt{1-(\frac{x}{1+\beta x})^2}}  d x d \beta 
+{\int^0_{-1}} P(\beta)\int^{\frac{1}{1-\beta}}_{0} \frac{ P(x=v/c_0)}{\sqrt{1-(\frac{x}{1+\beta x})^2}}  d x d \beta.
\label{12}
\end{eqnarray}
For example, if in Eq.(\ref{12}) we represent the probability $P(\beta)=\delta(\beta+0)+\delta(\beta-0)=\delta(\beta)$ ($\delta (x)=$ is the Dirac delta function), then we easily obtain the parameter $As=1$, i.e., the absence of asymmetry in the case of minimal energy, i.e., constant speed of light. The case of asymmetry is of interest when the corresponding probabilities are equally probable, i.e., $P(\beta)=1/2$ (since $\beta \in (-1,1)$), and $P(x)=const$. Then we can obtain the degree of asymmetry in the form $As=\frac{\pi+2}{\pi-2}\approx 4.5$, which means that matter exceeds antimatter by 4.5 times. Of course, by choosing different probability distributions, we can change the symmetry quite significantly. Also, if we assume that the probabilities are equal ($P(\beta)=const^{'}$ and $P(x)=const$), and the deviation from the minimum energy is the parameter $\beta$ (since the energy is minimal at $\beta=0$), and we denote this maximum deviation as $\beta_0$ (i.e. $\beta\in[-\beta_0,\beta_0]$), then it is easy to show that $As \geq 1$. Let's also represent the asymmetry parameter, often used in the literature, as $As_2=\frac{N_{+}-N_{-}}{N_{+}+N_{-}}$. This yields $As_2 \geq 1$, and the two asymmetry parameters under consideration will have a simple form
\begin{eqnarray}
As=\frac{b_0(\pi^2-4)-4\pi \sqrt{1-b^2_0}}{b_0(\pi^2+4)-4\pi},~~As_2=\frac{2}{\pi b_0}\left(1- \sqrt{1-b^2_0}\right) .
\label{13}
\end{eqnarray}
Let's also plot these functions, see Fig.\ref{fig_1}.
They are increasing functions with respect to $\beta_0$ with a maximum of $\beta_0=1$. In other words, if the probabilities of the appearance of particles with positive and negative energies are equal and constant under fluctuations of the energy minimum, then more particles with positive energy will always appear. It is interesting to note that this asymmetry does not arise due to a larger number of possible states in which a particle can exist (i.e., with an area of $\int d b dx$), since such an area is the same for $N_+$ and $N_-$. This asymmetry arises from the larger number of $N_+$ particles in the entire $\int d b dx$ region. If we assume that, as a result of certain interactions or the so-called "Big Bang," a deviation (even a very small fluctuation) of the minimum energy, that is, the $\beta$ parameter, from zero occurs, then matter will always predominate over antimatter in our world.
We should add that the deviation from the minimum energy at $\beta=0$ may arise due to quantum effects. In this case, the energy is not precisely defined and is related to the Heisenberg uncertainty principle, resulting in the energy deviating from its minimum, leading to an asymmetry in the Eq.{\ref{9}} under consideration.

\section{Conclusion}
Thus, in this paper it is shown that the speed of light is a constant at the minimum value of the particle energy. In general, space-time transformations are invariant at the speed of light $c(|v|)=\beta |v|+c_0$, where $|\beta| \leqslant 1$. A relationship between energy and momentum was also found, see Eq.(\ref{9}), which contains both positive and negative values of energy, in addition, Eq.(\ref{9}) is asymmetric. It was shown that quantum mechanics in the known representation can be realized only at $\beta=0$, i.e., at the minimum of energy. It was shown that due to the asymmetry of the relationship between energy and momentum, the amount of matter (solutions with positive energy) predominates over antimatter (solutions with negative energy) during fluctuations of the energy minimum. Certainly, the results obtained are interesting from a fundamental point of view, and the proof of the constancy of the speed of light presented here will serve as a starting point in the study of not only space and time, but also various areas of physics.

\normalsize

\section*{Disclosures}
The authors declare no conflicts of interest

\end{document}